\documentclass{iopart}
\usepackage{iopams}  

\newcommand{\Tb}{Tb$_{2}$Ti$_{2}$O$_{7}$}
\newcommand{\Is}{\ensuremath{\langle 111 \rangle}}

\begin{document}

\title[HFM2003]{The Spin Liquid State of
the Tb$_2$Ti$_2$O$_7$ Pyrochlore Antiferromagnet: A Puzzling State of Affairs}

\author{M. Enjalran$^{1,\dag a}$,  M.J.P. Gingras$^{1,2}$, 
Y.-J. Kao$^{1,\dag b}$, A. Del Maestro$^{1,\dag c}$, 
and H.R. Molavian$^{1}$
}

\address{$^{1}$ Department of Physics, University of Waterloo, 
Waterloo, Ontario, N2L 3G1, Canada }

\address{$^{2}$Canadian Institute for Advanced Research, 
180 Dundas Street West, Toronto, Ontario, M5G 1Z8, Canada }

\begin{abstract}
The pyrochlore antiferromagnet \Tb\ has proven
to be an enigma to experimentalists and theorists
working on frustrated magnetic systems.
The experimentally determined energy level structure suggests a
local \Is\ Ising antiferromagnet at
low temperatures, $T \stackrel{<}{\sim} 10$K.
An appropriate model then predicts 
a long-range ordered ${\mathbf Q}= {\mathbf 0}$ 
state below approximately $2$K. However, muon spin resonance ($\mu$SR)
experiments reveal a paramagnetic structure down to tens of milli-Kelvin.
The importance of fluctuations out of the ground state
effective Ising doublet has been recently understood,
for the measured paramagnetic correlations can
not be described without including the higher crystal field states.
However, these fluctuations treated within the random phase 
approximation (RPA) fail to account for the lack of ordering 
in this system below $2$K.
In this work, we briefly review the experimental evidence 
for the collective paramagnetic state of \Tb . The basic theoretical 
picture for this system is discussed, 
where results from classical spin models are used to motivate 
the investigation of quantum effects to lowest order via the RPA. 
Avenues for future experimental and theoretical work on \Tb\ 
are presented.  
\end{abstract}



%\maketitle

\section{Introduction}
\label{sec-intro}

Frustrated magnetism has been actively studied for 
decades,\cite{revs,rev-SI,lhuillier} 
and interest continues to grow. 
In models like the antiferromagnetic (AFM) $J_1$-$J_2$ model on a bipartite 
lattice, frustration results from a competition between two different
ground state spin structures, one favored by $J_1$ and the 
other by $J_2$.~\cite{chandra} 
The $J_1$-$J_2$ model has motivated many
interesting theoretical studies, but the tuning of 
energy scales to induce frustration 
is not easily achieved experimentally.
Another class of materials and models exhibit frustration at 
the level of nearest neighbor interactions via a 
competition imposed by the lattice geometry, creating 
geometric frustration.~\cite{sg-comment} 
Crystal geometries formed from unit triangles and 
AFM interactions are frustrated because the spins 
about a triangle can not arrange themselves so that
all pairwise interactions are satisfied 
(i.e., $-J_1{\mathbf S}_i\cdot{\mathbf S}_j$ for $J_1<0$ can
not be minimized for each pair $(i,j)$). Hence, the triangular 
and kagom\'e lattices 
are common frustrated geometries in $2d$. In three dimensions,
lattices of corner sharing triangles (garnet structures)
or tetrahedra (pyrochlore and spinel structures) are 
frustrated with nearest neighbor AFM interactions. 

The pyrochlore lattice has received increased attention recently 
because the $S=1/2$ AFM model is predicted to be a 
collective paramagnet, or spin liquid, in the ground 
state.~\cite{canals-prl,berg-hermele} 
A $3d$ spin liquid is unusual,~\cite{anderson-rvb}
but the role of lattice dimensionality appears to be reduced 
in the face of large quantum fluctuations
driven by frustration.~\cite{canals-prl}   
The stabilization of a 
non-collinear long-range ordered state
is expected 
when long-range interactions are included,~\cite{anderson-pyro} or 
when constraints arising from single-ion anisotropy are 
imposed.~\cite{111-moessner-prb} 
Magnetic insulating pyrochlores, general formula A$_2$B$_2$O$_7$,
where A is a rare earth ion 
(Ho$^{3+}$, Dy$^{3+}$, Tb$^{3+}$, Gd$^{3+}$), are a subset of frustrated
magnets that are at the focus of much experimental and
theoretical work.\cite{revs,rev-SI} This family
of materials displays long-range order,~\cite{gdtio} 
a novel ``ice like'' phase with residual entropy,
~\cite{rev-SI,1stSI,ramirez-nature} 
and collective paramagnetic behavior at milli-Kelvin 
temperatures.~\cite{tbtio-gardner1} 
The focus of our brief review is the candidate $3d$ spin 
liquid \Tb .~\cite{tbtio-gardner1} The many conflicting 
experimental results and theoretical predictions of this 
material are discussed, and the recent progress 
toward a general picture for \Tb\ is presented.  
  
\section{Experimental Picture of \Tb}
\label{sec-expt}

The pyrochlore \Tb\ has dominant AFM interactions 
as determined from dc magnetization measurements on a polycrystalline 
sample, $\theta_{\rm CW}\approx -20$K.~\cite{tbtio-gardner1}
Results from muon spin resonance ($\mu$SR) clearly indicate 
dynamic moments down to $70$mK despite the 
development of short-range correlations at approximately
$50$K, as seen in neutron scattering measurements.
~\cite{tbtio-gardner1,kanda-jps} 
Suppression of an ordering temperature in this material 
by two orders of magnitude 
(i.e., $|\theta_{\rm CW}|/70{\rm mK}= O(10^2)$) 
can be understood in the context of a nearest neighbor AFM.  
However, \Tb\ has a measured spin anisotropy gap 
of $\Delta \approx 18$K,~\cite{tbtio-gardner1,tbtio-gingras1} 
attributable to crystal field effects, 
as well as reasonably strong dipolar 
interactions,~\cite{tbtio-gingras1,dipSImodel1}
that such a large reduction in $T_c$ is quite puzzling.   
 
As a first step to unraveling the mystery of \Tb , one has
to determine the single ion properties of Tb$^{3+}$ in the dense
material. From experiments and calculations, Gingras {\it et al.} 
find that the Tb$^{3+}$ moment is approximately $5\mu_{\rm B}$, 
a substantial reduction from the free ion result 
$9.6\mu_{\rm B}$.~\cite{tbtio-gingras1}   
In addition, the crystal field effects on 
Tb$^{3+}$ ($S=3, L=3$ so $J=6$) yield an effective Ising 
ground state level structure 
with a quantization axis oriented along the local
\Is\ direction,~\cite{tbtio-gingras1,hotio-rosenkranz} 
with a gap to the next energy levels, an excited doublet, 
on the order of $\Delta \approx 18$K. 
This \Is\ anisotropy is significant since for 
AFM effective nearest neighbor interactions 
it greatly reduces the frustration from that of 
a Heisenberg like model.~\cite{111-moessner-prb} 
With a better grip on the dipolar strength and the crystal field 
levels, a revised Curie-Weiss temperature of $\theta_{\rm CW} \approx -14$K 
for the dipoles and exchange is calculated and an ordering temperature
of ~$1$K deduced.~\cite{tbtio-gingras1,dipSImodel1} 
Also, from theoretical work on a dipolar spin ice model (\Is\ Ising)
one is able to show that for the parameters of \Tb , 
a long-range ordered state with zero net moment about each tetrahedron 
(an all-in all-out state)
is expected at $T_c \approx 1.1$K.~\cite{dipSImodel1}
That experiments prove these predictions wrong is possibly due 
to the effects of long-range exchange beyond nearest-neighbor 
or quantum fluctuations.

With the single ion properties suggesting a \Is\ Ising dipolar
model for \Tb\ at $T<10$K, but the absence of any agreement between
this model and experiments on an ordering temperature, a 
re-evaluation of the model is necessary. Hence, the focus of the 
discussion shifts to the paramagnetic (PM) regime and the 
short-range correlations.  Early inelastic neutron data 
on a single crystal found evidence for the partial 
softening of the lowest energy modes 
about $[002]$ and $[220]$.~\cite{kanda-jps,tbtio-gardner2} 
Elastic scattering measurements on 
a single crystal of \Tb\ clearly display diffusive
scattering about these same regions in the $(hhl)$ plane with
the strongest correlations centered about 
$[002]$.~\cite{tbtio-gardner2,tbtio-yasui1} 
The qualitative features of
the structure factor ($S({\mathbf q})$) in the PM 
phase can be roughly fitted with 
a near neighbor AFM Heisenberg model,~\cite{tbtio-gardner2}
suggesting that the strict Ising anisotropy might
not hold for temperatures defining the PM  
regime of this material, i.e., $T \stackrel{>}{\sim} 0.5$K. 
Note, the PM $S({\mathbf q})$ is measured at 
$9$K in Gardner {\it et al.}~\cite{tbtio-gardner2} 
but at $0.4$K in Yasui {\it et al.}~\cite{tbtio-yasui1}
A relaxation of the \Is\ spin anisotropy in a long-range dipolar 
AFM pyrochlore model does yield good agreement with 
the PM $S({\mathbf q})$ in this material for temperatures
$T \stackrel{>}{\sim}|\theta_{\rm CW}|\stackrel{>}{\sim}10$K.
~\cite{enjalran,tbtio-yjkao} 
 
Recent data for \Tb\ in the low temperature regime 
suggests the existence of
a spin glass like state. Hysteresis is observed 
in the field cooled and zero field cooled neutron 
scattering data below $1.5$K.~\cite{tbtio-yasui1} 
Thermodynamic measurements of the static susceptibility also indicate 
spin glass physics, albeit at a much lower temperature, 
$70$mK.~\cite{corruccini-tbtio}.  

Hence, at present the general consensus is that \Tb\ remains
a collective paramagnet down to $T = 50$-$70$ mK at ambient
pressures. Under high pressure, $2-8$ GPa, 
the material orders at $T_c \approx 2$K,~\cite{mirebeau}  
but the spin structure is not the ${\mathbf Q}={\mathbf 0}$ 
all-in all-out structure predicted from
a \Is\ Ising model,~\cite{dipSImodel1,tbtio-yjkao}
or the ``nearby'' long-range ordered spin 
ice state.~\cite{SI-lro} 
An applied field can also induce an ordered state in this 
material.~\cite{tbtio-yasui2}
Field driven ordered states is a current topic in the spin
ice material Dy$_2$Ti$_2$O$_7$,\cite{magn-dytio}
but unlike the spin ice materials
moderate field strengths along high symmetry directions indicate
weaker spin anisotropy for \Tb .~\cite{tbtio-yasui1} 
We note that the gap between the ground state doublet and the 
first excited states in \Tb\ is an order of magnitude 
smaller than the gap measured
in the spin ice materials.~\cite{hotio-rosenkranz}
 
\section{Theoretical Interpretation of \Tb}
\label{sec-thry}

The Heisenberg AFM on the pyrochlore lattice
is defined by the simple Hamiltonian 
$H=-J_1\sum_{\langle (i,a),(j,b) \rangle}
{\mathbf S}_{i}^{a} \cdot {\mathbf S}_{j}^{b}$ ,
where $J_1<0$, $(i,a)$ describes the location of
a spin by its fcc lattice position ${\mathbf R}_i$ and 
tetrahedral basis point ${\mathbf r}^a$, and interactions
are among nearest neighbors only.  
The Hamiltonian $H$ represents a highly frustrated model 
that produces a spin liquid ground state for quantum 
spins (${\mathbf S}_i^a=1/2$) \cite{canals-prl,berg-hermele} 
or classical spins 
(${\mathbf S}_i^a=\hat{x}S_i^{a,x}+\hat{y}S_i^{a,y}+\hat{z}S_i^{a,z}$).
\cite{moessner-chalker-prl,moessner-chalker1,reimers1,reimers2}
In the classical model, an ordered state can be induced by 
perturbations arising from exchange interactions beyond 
nearest neighbor\cite{reimers1} 
or by constraints imposed by spin isotropy.~\cite{111-moessner-prb} 
If the single-ion anisotropy is along the local \Is\ direction
(i.e., ${\mathbf S}_{i}^{a} = \hat{z}^a \sigma_i^a$ where 
$\hat{z}^a$ is the local \Is\ direction and $\sigma_i^a= \pm 1$),
then it is the nearest neighbor ferromagnetic pyrochlore (nearest
neighbor spin ice) that is frustrated.~\cite{rev-SI,111-moessner-prb,1stSI} 
For rare earth magnets like the spin ice materials and \Tb ,
long-range dipolar interactions are a significant contribution to 
the Hamiltonian. Therefore, the basic model for rare earth pyrochlore 
insulators can be written,
\begin{equation}
\label{eq-Hdip}
\fl H = -J_1\sum_{\langle (i,a),(j,b) \rangle}
{\mathbf S}_{i}^{a} \cdot {\mathbf S}_{j}^{b} 
+ D_{\rm dd} \sum_{(i,a)>(j,b)}
\left ( \frac{{\mathbf S}_{i}^{a} \cdot {\mathbf S}_{j}^{b}}
{|{\mathbf R}_{ij}^{ab}|^3}
- \frac{3({\mathbf S}_{i}^{a} \cdot {\mathbf R}_{ij}^{ab})
({\mathbf S}_{j}^{b} \cdot {\mathbf R}_{ij}^{ab})}{|{\mathbf R}_{ij}^{ab}|^5} \right )
\ ,
\end{equation}
where $D_{\rm dd}=DR^3_{\rm nn}$ is the nearest neighbor 
dipole strength, $D=\mu_o \mu^2/4\pi R^3_{\rm nn}$, $\mu$ is the moment on the 
rare earth ion, $R_{\rm nn}$ is the nearest neighbor distance,  
and ${\mathbf R}_{ij}^{ab}$ is the vector separation between spins
${\mathbf S}_{i}^{a}$ and ${\mathbf S}_{j}^{b}$. If 
the spins in Eq.~\ref{eq-Hdip} are constrained
to their local \Is\ Ising axes, then  Eq.~\ref{eq-Hdip} 
represents the dipolar \Is\ Ising model (DIM) 
that successfully describes the specific heat 
\cite{dipSImodel1} and paramagnetic correlations 
\cite{hotio-spincorrel} of spin ice. When the DIM 
is simulated by a Monte Carlo algorithm employing nonlocal
loop moves, a first order transition to a spin ice ground 
state is found,~\cite{SI-lro} with an ordering wave vector 
and spin structure in agreement with mean-field 
calculations.~\cite{enjalran} 

The phase diagram of the DIM on the
pyrochlore lattice is generated by defining the sign and 
magnitude of $J_1$
and the strength of the 
dipole-dipole interaction D.~\cite{dipSImodel1}
The DIM exhibits two phases depending on the 
ratio $J_1/D$: 
For $J_1/D\stackrel{>}{\sim} -4.525$
a spin ice manifold results (macroscopic degeneracy), 
where two spins point in 
and two spins point out along their local \Is\ axes for
each unit tetrahedron. 
For $J_1/D < -4.525$,
a non-collinear long-range ordered state results,  
where all spins
about a unit tetrahedron point either in or out along their
local \Is\ axes, this ${\mathbf Q}={\mathbf 0}$ structure is replicated 
over the lattice. 
%new lines
We note that going from the PM to the ${\mathbf Q}={\mathbf 0}$ 
state occurs via a second order 
phase transition, but the path from the PM to spin ice 
manifold represents a cross over 
to a dynamically frozen state without long-range order. In the DIM, 
the temperature 
of the phase transition 
or cross over is signalled by the peak in the calculated specific heat,
whose position depends on the value of $J_1/D$.~\cite{dipSImodel1}
%end new lines
Hence, the important energy scale in this model is the 
effective nearest neighbor exchange, 
$J_{\rm eff}= J_1/3 + 5D/3$, where $1/3$ and 
$5/3$ are geometric factors arising from the inner product 
of the local \Is\ spin quantization axes. 
With $J_1/D \approx -5.5$ for \Tb ,
the prediction is a ${\mathbf Q}={\mathbf 0}$ 
state at approximately $1$K.~\cite{tbtio-gingras1,dipSImodel1} 
The subtle point about \Tb\ in the DIM is that
it sits near the boundary between the two
phases, making it susceptible to fine tuning of 
the model parameters (exchange or dipolar energy scales
and single ion crystal field state wave functions). 
It is conceivable that \Tb\ might sit close to a line of disorder 
in the phase diagram where fluctuations dominate and 
a tendency to order is suppressed. The simulation of the 
DIM near the phase boundary presents some numerical 
challenges and is not well understood.~\cite{SI-lro}

As mentioned in Section \ref{sec-expt}, \Tb\ 
does not order at $T \approx 1$K, and suggestions from 
the PM $S(\mathbf q)$ results 
\cite{tbtio-gardner2,tbtio-yasui1} that the
\Is\ Ising constraint should be relaxed in this case
provides a hint of where to venture next. 
In considering the symmetry 
of the DIM, one can show that no strict
\Is\ Ising model on the pyrochlore lattice 
can reproduce the short-range correlations observed in 
experiments, but a model in which the rotational symmetry is
partially restored is found to admit a solution compatible
with experiments.~\cite{enjalran} Mean-field and Monte Carlo
calculations make these claims more quantitative.~\cite{enjalran} 
Hence, in the PM regime the physics of \Tb\ 
can not be ascribed to the same \Is\ Ising interpretation 
of the moments that works so well for the spin ice materials.     

A pitfall with a classical Heisenberg description 
for \Tb\ is that complete restoration of spin isotropy does 
not agree with experiments and calculations of 
the single ion properties of the 
Tb$^{3+}$.~\cite{tbtio-gingras1,hotio-rosenkranz}
The g-tensor of Tb$^{3+}$ in \Tb\ exhibits \Is\ like anisotropy below 
$10$K and a ground state
doublet structure that is separated from a first excited
doublet by about $18$K, which is on the order of the Curie-Weiss 
temperature for exchange and dipolar interactions 
($\theta_{\rm CW}=-14$K).~\cite{tbtio-gingras1} 
For the spin orbit coupled
Tb$^{3+}$ moments ($J=6$), the angular 
momentum for the two lowest lying doublets is predominately 
$|M_J=\pm 4 \rangle$ for ground
state and $|M_J=\pm 5 \rangle$ for the first
excited state.~\cite{tbtio-gingras1} 
Hence, transverse fluctuations between these
two lowest lying doublets are allowed and 
non-negligible at the temperatures of the
$S({\mathbf q})$ experiments, 
$5$-$10$K.~\cite{tbtio-gardner2,tbtio-yjkao}

Applying a random phase approximation (RPA) to
a model with exchange and long-range dipoles, the
fluctuations between the lowest lying 
crystal field states can be studied, i.e., 
the \Is\ Ising limit is recovered when
transitions between states $|\psi_0 \rangle$ (ground state) 
and $|\psi_1 \rangle$ (first excited state)
have a vanishing matrix element, 
$\langle \psi_1|S^\pm|\psi_0 \rangle =0$.
Using $|\psi_0 \rangle = |M_J=\pm 4\rangle$ 
and $|\psi_1 \rangle = |M_J=\pm 5>$, the 
RPA~\cite{tbtio-yjkao} gives excellent semi-quantitative agreement 
with the PM $S({\mathbf q})$ and the energy dispersion 
between the low lying magnetic states of the 
neutron experiments.~\cite{tbtio-gardner2}
However, the RPA still predicts an ordered state at 
$T_c \approx 2$K. Hence, the RPA demonstrates the sensitivity of the 
PM correlations observed in neutron scattering
to quantum fluctuations in this unusual magnetic 
system ($|\psi_0 \rangle = |M_J=\pm 4\rangle$ and 
$|\psi_1 \rangle = |M_J=\pm 5>$), but it
still gives a $T_c$ of approximately $2$K to the 
above all-in all-out ${\mathbf Q}={\mathbf 0}$ state.

\section{Avenues for Future Studies and Conclusions}

%Future directions
In order to understand the physics of \Tb\ below 
$2$K, we need to go beyond the first order approximation
of the RPA and consider quantum effects more rigorously. 
A reasonable first step would be to employ a more detailed 
description of the crystal field states of the Tb$^{3+}$ ion, 
and then form an effective low energy Hamiltonian from the more 
complicated model following a procedure similar to that used
in the derivation of the t-J or Heisenberg models from the
Hubbard model. 
Experimentally, low temperature neutron scattering (elastic and 
inelastic) measurements on single crystals are needed to probe
the fluctuations between crystal field 
states.~\cite{tbtio-gardner1,kanda-jps,tbtio-gardner2} 
Exploring these same properties but at hydrostatic 
pressures below $2$GPa 
could provide a valuable link to the high pressure ordered
state of \Tb .~\cite{mirebeau} 

%Conclusions
In conclusion, there is strong evidence that 
\Tb\ can not be described by a classical \Is\ Ising Hamiltonian
similar to the successful model of the spin ice 
materials.~\cite{rev-SI} 
Spin fluctuations must be incorporated into any model of \Tb\ 
in order to capture the PM correlations, and a simple description 
of quantum fluctuations via the RPA is sufficient to 
achieve good agreement with experiments. 
However, to study the lack of an ordering transition in \Tb , 
a more careful treatment of the quantum fluctuations is essential.
Finally, we come back to the \Is\ Ising dipolar model phase
diagram and note that \Tb\ is very close to the AFM/spin ice phase
boundary.~\cite{dipSImodel1} There may be issues with the quantum fluctuations,
long-range exchange, microscopic disorder at very low temperatures, 
or tuning of material or model parameters that places \Tb\ on 
a line or region of disordered semi-classial ground states.   

We thank R.G. Melko, S. Bramwell, B. Gaulin, J. Gardner, H. Fukazawa, 
R. Higashinaka, and Y. Maeno for useful discussions. M.G. acknowledges
financial support from NSERC of Canada, the Canada Research
Chair Program, Research Corporation, and the Province of Ontario.

%\begin{thebibliography}{99}


\Bibliography{99}
\bibitem[\dag]{note1} Addresses after September 1: 
$^a$ Department of Physics, Southern Connecticut State University, 
501 Crescent St., New Haven, Connecticut, 06515; $^b$
Department of Physics, University of Toronto, 60 St. George St.,
Toronto, M5S 1A7, Ontario, Canada; $^c$ Department of 
Physics, Sloane Physics Lab, Yale University,
217 Prospect St., New Haven, Connecticut, 06520-8120.

%\bibitem[\ddag]{note2} Address after September 1:
%Department of Physics, University of Toronto, 60 St. George St.,
%Toronto, M5S 1A7, Ontario, Canada.

%\bibitem[\S]{note3} Address after September 1:
%Department of Physics, Sloane Physics Lab,
%Yale University,
%217 Prospect St., New Haven, Connecticut, 06520-8120.
 
%\bibitem{laughlin-coleman} R. B. Laughlin and D. Pines,
%Proc. Natl. Acad. Sci. U.S.A {\bf 97}, 28 (2000); 
%P. Coleman, Ann. Henri Poincar\'{e} {\bf 4}, 1 (2003),
%cond-mat/0307004.

\bibitem{revs} A. P. Ramirez,
Annu. Rev. Mater. Sci. {\bf 24}, 453 (1994);
J. E. Greedan, J. Mater. Chem. {\bf 11}, 37 (2001); 
{\textit Highly Frustrated Magnetism 2000, Proceedings}, 
Can. J. Phys. {\bf 79}, 2001.

\bibitem{rev-SI} S. T. Bramwell and M. J. P. Gingras,
Science {\bf 294}, 1495 (2001).

\bibitem{lhuillier} C. Lhuiller and G. Misguich, cond-mat/0109146.
Lectures presented at the Cargese summer school on ``Trends in 
high-magnetic field science'', May 2001.

\bibitem{chandra} P. Chandra and B. Doucot,
Phys. Rev. B {\bf 38} 9335 (1988).

\bibitem{sg-comment}
Frustration also results if there is distribution of 
interaction strengths between moments, 
such systems (spin glasses) are not considered here.

\bibitem{canals-prl} B. Canals and C. Lacroix,
Phys. Rev. Lett. {\bf 80}, 2933 (1998).

\bibitem{berg-hermele} E. Berg, E. Altman, and A. Auerbach,
cond-mat/0206384 (unpublished); 
M. Hermele, M. P. A. Fisher, and L. Balents,
cond-mat/0305401 (unpublished). 

\bibitem{anderson-rvb} P. W. Anderson, B. Halperin, and C. M. Varma,
Philos. Mag. {\bf 25}, 1 (1972).

\bibitem{anderson-pyro} P. W. Anderson,
Phys. Rev. {\bf 102}, 1008 (1956).

\bibitem{111-moessner-prb} R. Moessner,
Phys. Rev. B {\bf 57}, R5587 (1998); S. T. Bramwell, M. J. P. Gingras, 
and J. N. Reimers, J. Appl. Phys. {\bf 75}, 5523 (1994).

\bibitem{gdtio} N. P. Raju, M. Dion, M. J. P Gingras, T. E. Mason,
and J. E. Greedan, Phys. Rev. B {\bf 59}, 14489 (1999); 
S. E. Palmer and J. T. Chalker, Phys. Rev. B {\bf 62}, 488 (2000);
J. D. M. Champion, A. S. Wills, T. Fennell, 
S. T. Bramwell, J. S. Gardner, and M. A. Green,
Phys. Rev. B {\bf 64}, 140407 (2001).

\bibitem{1stSI} M. J. Harris, S. T. Bramwell, D. F. McMorrow, T. Zeiske, and 
K. W. Godfrey, Phys. Rev. Lett. {\bf 79}, 2554 (1997).

\bibitem{ramirez-nature} A. P. Ramirez, A. Hayashi, R. J. Cava, R. Siddharthan,
and B. S. Shastry, Nature {\bf 399}, 333 (1999). 

\bibitem{tbtio-gardner1} J. S. Gardner, S. R. Dunsiger, B. D. Gaulin, 
M. J. P. Gingras, J. E. Greedan, R. F. Kiefl, M. D. Lumsden, W. A. MacFarlane,
N. P. Raju, J. E. Sonier, I. Swainson, and Z. Tun,
Phys. Rev. Lett. {\bf 82}, 1012 (1999).

\bibitem{kanda-jps} M. Kanada, Y. Yasui, M. Ito, H. Harashina, M. Sato, 
H. Okumura, and K. Kakurai, J. Phys. Soc. Jpn. {\bf 68}, 3802 (1999). 

\bibitem{tbtio-gingras1} M. J. P. Gingras, B. C. den Hertog, M. Faucher, 
J. S. Gardner, S. R. Dunsiger, L. J. Chang, B. D. Gaulin, N. P. Raju, 
and J. E. Greedan, Phys. Rev. B {\bf 62}, 6496 (2000).

\bibitem{dipSImodel1} B. C. den Hertog and M. J. P. Gingras,
Phys. Rev. Lett. {\bf 84}, 3430 (2000).

\bibitem{hotio-rosenkranz} S. Rosenkranz, A. P. Ramirez, A. Hayashi, 
R. J. Cava, R. Siddharthan, and B. S. Shastry,
J. Appl. Phys. {\bf 87}, 5914 (2000).

\bibitem{tbtio-gardner2} J. S. Gardner, B. D. Gaulin, A. J. Berlinsky, 
P. Waldron, S. R. Dunsiger, N. P. Raju, and J. E. Greedan,
Phys. Rev. B {\bf 64}, 224416 (2001).

\bibitem{tbtio-yasui1} Y. Yasui, M. Kanada, M. Ito, H. Harashina, M. Sato,
H. Okumura, K. Kakurai, and H. Kadowaki,
J. Phys. Soc. Jpn. {\bf 71}, 599 (2002).

\bibitem{enjalran} M. Enjalran and M. J. P. Gingras,
cond-mat/0307151 (unpublished).

\bibitem{tbtio-yjkao} Y.-J. Kao, M. Enjalran, A. Del Maestro, 
H. R. Molavian, and M. J. P. Gingras, cond-mat/0207270. To appear in
Phys. Rev. B (BR).

\bibitem{corruccini-tbtio} G. Luo, S. T. Hess, and L. R. Corruccini,
Phys. Lett. A {\bf 291}, 306 (2001).

\bibitem{mirebeau} I. Mirebeau, I. N. Goncharenko, P. Cadavez-Perez, 
S. T. Bramwell, M. J. P. Gingras, and J. S. Gardner,
Nature {\bf 420}, 54 (2002).

\bibitem{SI-lro} R. G. Melko, B. C. den Hertog, and M. J. P. Gingras,
Phys. Rev. Lett. {\bf 87}, 067203 (2001); R. G. Melko, thesis, 2001, 
University of Waterloo, Ontario, Canada. 

\bibitem{tbtio-yasui2} Y. Yasui, M. Kanada, M. Ito, H. Harashina, M. Sato, 
H. Okumura, and K. Kakurai, J. Phys. and Chem. Solids {\bf 62}, 343 (2001).

\bibitem{magn-dytio} H. Fukazawa, R. G. Melko, R. Higashinaka, 
Y. Maeno, and M. J. P. Gingras, Phys. Rev. B {\bf 65}, 054410 (2002);
R. Higashinaka, H. Fukazawa, and Y. Maeno, 
cond-mat/0305529 (unpublished); Z. Hiroi, K. Matsuhira, and M. Ogata, 
cond-mat/0306240 (unpublished). 

\bibitem{moessner-chalker-prl} R. Moessner and J. T. Chalker,
Phys. Rev. Lett. {\bf 80}, 2929 (1998).

\bibitem{moessner-chalker1} R. Moessner and J. T. Chalker,
Phys. Rev. B {\bf 58}, 12049 (1998).

\bibitem{reimers1} J. N. Reimers, A. J. Berlinsky, and A.-C. Shi,
Phys. Rev. B {\bf 43}, 865 (1991). 

\bibitem{reimers2} J. N. Reimers, 
Phys. Rev. B {\bf 46}, 193 (1992).

%\bibitem{canals-cjp} B. Canals and D. A. Garanin,
%Can. J. Phys. {\bf 79}, 1323 (2001).

\bibitem{hotio-spincorrel} S. T. Bramwell, M. J. Harris, B. C. den Hertog, 
M. J. P. Gingras, J. S. Gardner, D. F. McMorrow, A. R. Wildes, 
A. L. Cornelius, J. D. M. Champion, R. G. Melko, and T. Fennell,
Phys. Rev. Lett. {\bf 87}, 047205 (2001).


%\end{thebibliography}
\endbib
\end{document}